\documentclass[12pt]{article}
\usepackage{graphicx}
\usepackage{type1cm, amsmath, amssymb}

\newcommand{\nn}{\nonumber}

\newcommand{\Ncal}{\mathcal{N}}
\newcommand{\Fcal}{\mathcal{F}}
\newcommand{\del}{\partial}
\newcommand{\deldel}[2]{\frac{\del #1}{\del #2}}
\newcommand{\vp}{\varphi}
\DeclareMathOperator{\Tr}{{\rm Tr}}
\DeclareMathOperator{\tr}{{\rm tr}}

\newcommand{\rap}[2]
{\setbox1=\hbox{#1}%
\setbox2=\hbox to\wd1{\hss #2\hss}%
\mbox{\rlap{\box1}\box2}}
\newcommand{\sla}[1]{\rap{$#1$}{/}}

\newcommand{\thet}{\theta_{k}}
\newcommand{\uk}{u_k}
\newcommand{\sigmac}{\check{\sigma}}
\newcommand{\sigc}{\sigma_{C}}
\newcommand{\sigmah}{\hat{\sigma}}

\newcommand{\mmev}[1]{\left\langle #1\right\rangle_{\text{mm}}}
\newcommand{\mcl}{m^{(\text{cl})}}
\newcommand{\Scal}{{\cal S}}
\newcommand{\bulk}{{\text{bulk}}}
\newcommand{\bdy}{{\text{bdy}}}
\newcommand{\p}{$\bigcirc$}
\newcommand{\x}{$\times$}
\newcommand{\Lcal}{{\cal L}}
\newcommand{\chic}{\check{\chi}}
\newcommand{\chih}{\hat{\chi}}
\newcommand{\nablac}{\overset{\circ}{\nabla}}

\numberwithin{equation}{section}

\begin{document}

\thispagestyle{empty}

\begin{flushright}
 \begin{tabular}{l}
 {\tt hep-th/0603208}\\
 IHES/P/06/18
 \end{tabular}
\end{flushright}

 \vfill
 \begin{center}
{\LARGE
 \centerline{Wilson Loops of Anti-symmetric 
  Representation and D5-branes} 
 \vskip 3mm
 \centerline{ }
}

 \vskip 2.0 truecm
\noindent{ \large  Satoshi Yamaguchi} \\
{\sf yamaguch@ihes.fr}
\bigskip

 \vskip .6 truecm
 {
 {\it 
IHES, Le Bois-Marie, 35, route de Chartres\\ 
F-91440 Bures-sur-Yvette,
FRANCE
} 
 }
 \vskip .4 truecm

 \end{center}

 \vfill
\vskip 0.5 truecm

\begin{abstract}
We use a D5-brane with electric flux in $AdS_5\times S^5$ background to calculate the circular Wilson loop of anti-symmetric representation in $N=4$ super Yang-Mills theory in 4 dimensions. The result agrees with the Gaussian matrix model calculation.
\end{abstract}
\vfill
\vskip 0.5 truecm

\newpage

\section{Introduction and summary}
The expectation values of Wilson loops are some of the most important kinds of observables in a gauge theory. In particular, Wilson loops are interesting operators in AdS/CFT correspondence since they are calculable in AdS side using strings \cite{Rey:1998ik,Maldacena:1998im}.

In this paper, we are interested in the Wilson loop operators of the following kind in $\Ncal=4$ Yang-Mills theory.
\begin{align}
 \Tr_{R}\left[P\exp\left(i\oint_{C} ds (A_{\mu}\dot{x}^{\mu}+\vp_4 
 |\dot{x}|)\right)\right], \label{wilson-loop}
\end{align}
where $\vp_4$ is a scalar field in $\Ncal=4$ Yang-Mills theory, $R$ is a representation of the gauge group U$(N)$. This is a Lorentzian version of Wilson line, and we mainly consider the straight path $C$ in Lorentzian case. When one performs Wick-rotation and some conformal transformation, a circular Wilson loop is obtained.
\begin{align}
 W_{R}(C)=\Tr_{R}\left[P\exp\left(\oint_{C} ds (iA_{\mu}\dot{x}^{\mu}+\vp_4 
 |\dot{x}|)\right)\right], \label{circular-wilson-loop}
\end{align}
where $C$ is a circle in Euclidean 4-dimensional space. The expectation value of this operator depends on the representation $R$.

In the leading order of the calculation by fundamental string in AdS spacetime, namely large $N$ and large 't Hooft coupling $\lambda$ limit, we can hardly see the detail of this trace structure $R$. In this limit, the expectation value is expressed as $\exp(k\sqrt{\lambda})$ where $k$ is the number of boxes when $R$ is expressed by a Young diagram. This is because it is hard to distinguish the winding and overwrapping of the strings; we can just see the string charge $k$.

The approach using D3-brane with electric flux developed in \cite{Rey:1998ik} has been found to be useful to see this trace structure \cite{Drukker:2005kx}. The D3-brane with induced metric $AdS_2\times S^2$ corresponds to the circular Wilson loop operator $\tr[U^k]$, where $U=P\exp\left(\oint_{C} ds (iA_{\mu}\dot{x}^{\mu}+\vp_4 |\dot{x}|)\right)$ and $\tr$ denotes the trace in the fundamental representation $\Tr_{\square}$. This operator can be expressed in terms of a certain combination of the representation with $k$ boxes. The authors of \cite{Drukker:2005kx} have calculated the on-shell action of $AdS_2\times S^2$ D3-brane with $k$ unit of electric flux. They have shown that this on-shell action reproduce the VEV of the Wilson loop, which can be calculated by the Gaussian matrix model, in all order in $k\sqrt{\lambda}/N$ and leading order in $1/\lambda$. 

This kind of D3-brane for the Wilson loop is an analogue of a giant graviton for the half BPS local operator. Especially, $AdS_2\times S^2$ D3-brane is an analogue of the giant graviton wrapped on a $S^3$ in the $AdS_5$ \cite{Grisaru:2000zn,Hashimoto:2000zp}. There is also an analogue of the giant graviton wrapped on $S^3$ in the $S^5$ \cite{McGreevy:2000cw} for the Wilson line. This is a D5-brane wrapped on $S^4$ in the $S^5$ \cite{Pawelczyk:2000hy,Camino:2001at}.

In this paper, we consider this D5-brane. There is a BPS configuration of a single D5-brane with electric flux whose induced metric is $AdS_2\times S^4$. We will show that this D5-brane preserves the same half of the supersymmetry as the $AdS_2\times S^2$ D3-brane. This fact indicate that this D5-brane corresponds to the Wilson loop operator of \eqref{circular-wilson-loop} with some trace structure $R$.

Actually, we claim that this $AdS_2\times S^4$ D5-brane corresponds to the Wilson loop of \eqref{circular-wilson-loop} with {\em anti-symmetric tensor product of fundamental representation}. We will explain in this section the reason for this identification from two points of view. One is the bubbling geometry \cite{Yamaguchi:2006te}, which is a Wilson line version of LLM bubbling geometry \cite{Lin:2004nb}. The other is the D3-D5-F1 brane system.

First, let us explain from the bubbling geometry point of view. It is indicated in \cite{Yamaguchi:2006te} that the supergravity background which corresponds to a Wilson line is expressed as an $AdS_2\times S^2\times S^4$ fibration over 2-dimensional surface with boundary. At the boundary, the fiber becomes singular and either $S^2$ or $S^4$ shrinks. If we paint by black the points where $S^2$ shrinks, and by white where $S^4$ shrinks, then we will obtain one dimensional black and white pattern. This pattern seems to correspond to the eigenvalue distribution of the Gaussian matrix model, which is used to calculate the expectation value of the Wilson loops \cite{Erickson:2000af,Drukker:2000rr}.

For example the $AdS_5\times S^5$ can be expressed by the pattern with one black segment around the center. In this picture, the $AdS_2\times S^2$ D3-brane of \cite{Drukker:2005kx} can be expressed by a black point on a white part. This kind of eigenvalue distribution can be obtained by the insertion of the operator $\tr[e^{kM}]$, where k is an integer and $M$ is the matrix of the Gaussian matrix model. This is consistent with the fact that $AdS_2\times S^2$ D3-brane corresponds to the Wilson loop $\tr[U^k]$.

\begin{figure}
\begin{center}
  \includegraphics[width=10cm]{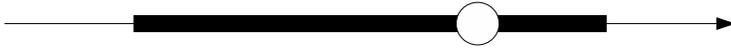}
\end{center}
\caption{The eigenvalue distribution with a hole or
 $AdS_5\times S^5$ geometry with an $AdS_2\times S^4$ D5-brane.}
\label{fig-hole}
\end{figure}

\begin{figure}
\begin{center}
  \includegraphics[width=10cm]{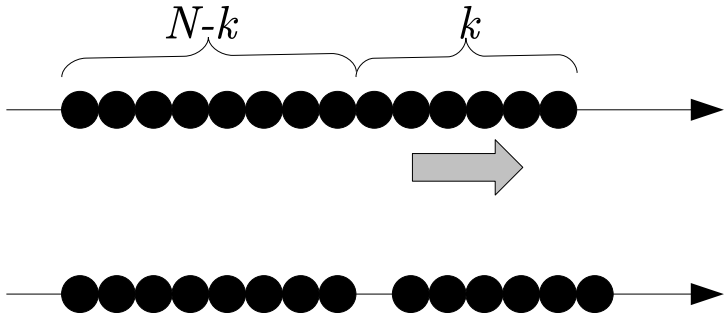}
\end{center}
\caption{The picture of eigenvalue distribution and external force. The upper figure expresses the eigenvalue distribution of the vacuum. When we add small external constant force to the first $k$ eigenvalues, these eigenvalues move a little to the right and there appears a ``hole'' next to the $k$-th eigenvalue.}
\label{fig-move}
\end{figure}

On the other hand, an $AdS_2\times S^4$ D5-brane looks like a white point on the black segment as shown in figure \ref{fig-hole}. In the matrix model, this looks like a hole in the eigenvalues. This kind of eigenvalue distribution can be obtained by the insertion of the operator $\Tr_{A_k}[e^{M}]$ where $A_k$ is the rank $k$ anti-symmetric representation. This is because of the following reason. When one insert this operator, $k$ eigenvalues out of $N$ feel a unit constant external force. This force moves these $k$ eigenvalues to the right. Then there appears a gap next to the $k$-th biggest eigenvalue as shown in figure \ref{fig-move}. This gap looks like a hole.

Next let us explain from the original D3-D5-F1 system point of view. We are considering the $N$ D3-branes, one D5-brane and $k$ fundamental strings between D3 and D5-brane, as shown in table \ref{tbl}. This configuration actually preserves 8 SUSY instead of 4 SUSY. This means adding D5-brane to D3-F1 system do not break father supersymmetry.
\begin{table}
 \begin{center}
  \begin{tabular}{|c||c|c|c|c|c|c|c|c|c|c|}\hline
   &0  &1  &2  &3  &4  &5  &6  &7  &8  &9 \\ \hline\hline
 D3&\p &\p &\p &\p &\x &\x &\x &\x &\x &\x \\ \hline
 D5&\p &\x &\x &\x &\p &\p &\p &\p &\p &\x \\ \hline
 F1&\p &\x &\x &\x &\x &\x &\x &\x &\x &\p \\ \hline
  \end{tabular}
 \end{center}
 \caption{The configuration of branes. Here {\p} denotes the direction parallel to the brane, and {\x} denotes the one perpendicular to the brane.}
\label{tbl}
\end{table}

\begin{figure}
 \begin{center}
  \includegraphics[width=10cm]{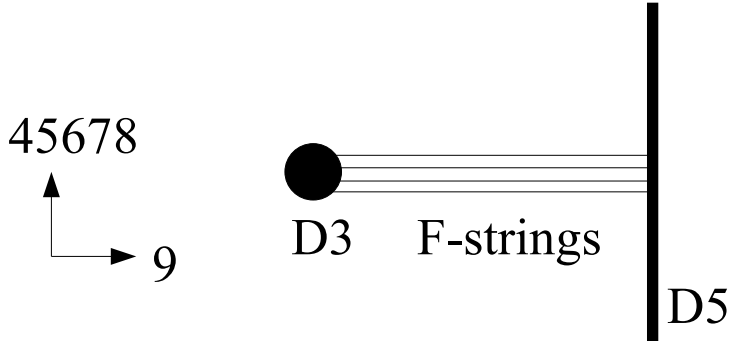}
 \end{center}
\caption{The picture of the branes. There are $N$ D3-branes, one D5-brane and $k$ fundamental strings stretched between the D3-branes and the D5-brane.}
\end{figure}
The fundamental string between D3 and D5 in this system behaves as a fermion because of the following reason\footnote{This is the same mechanism as the fermionic nature of the strings of baryons \cite{Witten:1998xy}.}. Let us consider this string in NSR formalism. Since the worldsheet theory has 8 Dirichlet-Neumann directions, the vacuum energy in the NS sector is greater than 0. Therefore the zero energy states are in R sector. We only have two fermion zero modes $\psi^{0}_{0},\psi^{9}_{0}$. The representation of these zero modes is two dimensional. The GSO projection leaves only one zero energy state. As a result, the only label is the Chern-Paton factor at the end of D3-brane side. These ground states are in R sector, so they behave as fermions, and the Chern-Paton indices are anti-symmetrized. As a result, this stack of $k$ strings looks like a particle in anti-symmetric representation of the gauge group U($N$) in the D3-branes.

In this configuration, we can express the $N$ D3-branes by the supergravity solution. The D5-brane and the fundamental strings are expressed by the spike solution of the D5-brane worldvolume theory \cite{Callan:1997kz,Gibbons:1997xz}. In the near horizon limit of the D3-branes, this D5-brane solution has induced metric of $AdS_2\times S^4$.

Let us summarize the result of this paper here. As already mentioned, it is shown that the $AdS_2\times S^4$ D5-brane preserves the same half of the supersymmetry as the $AdS_2\times S^2$ D3-branes. We also calculate the on-shell action of this D5-brane with careful treatment of the boundary terms. The result is
\begin{align}
 S_{\text{on-shell}}=-\sqrt{\lambda}\frac{2N}{3\pi}\sin^3\thet,
\end{align}
where $\thet$ is related to $k$ by
\begin{align}
 k=\frac{2N}{\pi}\left(\frac12\thet-\frac14 \sin 2\thet\right).
\end{align}
We compare this result to the matrix model calculation in a semi-classical approximation. They turn out to be completely the same in this level of approximation.

The construction of this paper is as follows. In section \ref{sec-susy}, we check the supersymmetry of $AdS_2\times S^4$ D5-brane. In section \ref{sec-circ}, we treat circular Wilson loop and compare the D5-brane calculation and matrix model calculation. 

Note: While we were finishing this paper, we found an interesting paper \cite{Hartnoll:2006hr} appeared in arXiv. The authors of \cite{Hartnoll:2006hr} use a D5-brane to calculate the Polyakov loop in thermal theory. Among their result, the ``constant solution'' seem to be essentially the same as the D5-brane solution of ours. In our paper, we identify this D5-brane solution as the Wilson loop of anti-symmetric representation. We compare this D5-brane calculation and the matrix model calculation in our paper.

\section{Supersymmetry of AdS D-branes}
\label{sec-susy}
In this section, we study the supersymmetry of the $AdS_2\times S^4$ D5-brane. We show that this D5-brane preserves just the same half of the supersymmetry as an $AdS_2\times S^2$ D3-brane. In this section, we work in Lorentzian signature.

\subsection{Supersymmetry of $AdS_5\times S^5$ spacetime}

It is convenient to express the $AdS_5\times S^5$ background as the fibration of 2-dimensional space with $AdS_2\times S^2\times S^4$ fiber 
\cite{Yamaguchi:2006te}. 
\begin{align}
 &ds^2=L^2\left[\cosh^2 u \, d\check{\Omega}_2^2+du^2+d\theta^2
   +\sinh^2 u \, d \hat{\Omega}_2^2+\sin^2\theta \, d\Omega_4^2\right],\label{metric1}\\
 &G_5=\frac{4}{L}\left[E^2 E^0 E^1 E^4 E^5+E^3 E^6 E^7 E^8 E^9
\right],\\
&u\ge 0,\qquad 0\le \theta\le \pi,
\end{align}
where $d\check{\Omega}_2^2,d \hat{\Omega}_2^2,d\Omega_4^2$ are the metrics
of unit $AdS_2$, $S^2$, $S^4$ respectively. The vielbein of the metric 
\eqref{metric1} are denoted by $E^{M},\ 
M=0,1,\dots,9$ and the vielbein of unit $AdS_2$ by $e^0,e^1$, unit 
$S^2$ by $e^4,e^5$, and unit $S^4$ by $e^{6},\dots , e^{9}$. With this notation, 
the vielbein of the fiber part of the metric \eqref{metric1} are expressed as $E^0=L \cosh u  e^{0},\ E^1=L \cosh u e^{1}$, and $E^4=L\sinh u e^{4},\ 
E^5=L\sinh u e^{5}$, and $E^6=L\sin\theta e^{6},\dots , E^9=L\sin\theta e^{9}$.
The vielbein of the base space part are expressed as $E^2=Ldu,\ E^3=Ld\theta$. In these equations $L$ is the radius of $AdS_5$ and $S^5$. It is expressed as $L=(\alpha'^2 4\pi g_s N)^{1/4}$. Here $g_s$ is the string coupling constant, and $N$ is the number of D3-branes. The potential $C_4$ for five form, which satisfies $dC_4=G_5$, can be written as
\begin{align}
 C_4=4L^4\left[
\left(-\frac{1}{8}u+\frac{1}{32}\sinh 4u\right)e^{0}e^{1}e^{4}e^{5}
+\left(\frac{3}{8}\theta-\frac{1}{4}\sin2\theta+\frac{1}{32}\sin4\theta\right)
e^{6}e^{7}e^{8}e^{9}
\right].\label{c4}
\end{align}

In order to investigate the supersymmetry preserved by D-branes, we need the explicit form of the Killing spinor in this spacetime. We can expand the 10-dimensional IIB type spinor as 
$\eta=\sum_{abcIJK}\epsilon_{abcIJK}\otimes
\chic_{a}^{I}\otimes \chih_{b}^{J}\otimes \chi_{c}^{K}$. The detail of this expansion is written in appendix \ref{app1}. With these notations, the supersymmetry condition can be written as
\begin{align}
 -i\mu_1\nu_3\lambda_3\epsilon
=\left[-\sinh u\Gamma^{2} +\cosh u \mu_3\nu_3\tau_2 \Gamma^2\right]\epsilon,\label{susycond1}\\
 \nu_1\lambda_3\epsilon
=\left[\cosh u \Gamma^{2}-\sinh u \mu_3\nu_3\tau_2 \Gamma^2\right]\epsilon,\\
 \lambda_1\epsilon
=\left[\cos \theta \Gamma^3 + \sin \theta \mu_3\nu_3\tau_2 \Gamma^{2}\right]\epsilon,\label{susycond3}\\
 \del_{u}\epsilon=\frac12 \mu_3\nu_3\tau_2 \epsilon,\qquad 
 \del_{\theta}\epsilon=\frac12 \mu_3\nu_3\tau_2\Gamma_{23} 
 \epsilon.\label{differential-equation}
\end{align}

These equations can be solved by the usual manner. First, \eqref{differential-equation} determines the coordinate dependence of $\epsilon$.
\begin{align}
 \epsilon=\exp\left(\frac12 \mu_3\nu_3\tau_2 u-\frac{i}{2}\lambda_3\tau_2\theta\right)\zeta,\label{epsilon}
\end{align}
where $\zeta$ is a constant spinor. Next, eqs.\eqref{susycond1}-\eqref{susycond3} imply some projection conditions on $\zeta$. These conditions are written as
\begin{align}
 &\mu_2\lambda_3\tau_2\Gamma^2\zeta=\zeta,\qquad
 \nu_1\lambda_3\Gamma^{2}\zeta=\zeta,\qquad
 \lambda_1\Gamma^3\zeta=\zeta. \label{proj1}
\end{align}
We should also assign the ten dimensional Weyl condition to this spinor. That condition can be expressed in our $\epsilon$ like
\begin{align}
 \Gamma_{23}\zeta=-i\mu_3\nu_3\lambda_3\zeta.\label{weyl-proj}
\end{align}
The number of independent components can be counted as follows. Originally each $\epsilon_{abcIJK}$ is a pair of 2-component spinors, so it has 4 (complex) components. Each of $a,b,c$ takes values $+1,-1$, and $I=1,2,\ J=1,2,\ K=1,2,3,4$. In summary, originally $\epsilon$ has $2^9$ complex components. Then there are four projection conditions, each of which project out half of the components. Then the independent complex components are $2^5=32$. Finally the Majorana condition relates $J=1$ components and complex conjugation of $J=2$ components. This leaves 32 real components independent.
\subsection{Supersymmetry of the $AdS_2\times S^2$ D3-branes}
Let us first consider here the supersymmetry preserved by an $AdS_2\times S^2$ D3-brane of \cite{Drukker:2005kx} in order to compare it to the one preserved by a D5-brane. In general, the supersymmetry preserved by branes can be examined by the kappa symmetry projection  $0=(1+\Gamma)\xi$ as in \cite{Cederwall:1996pv,Aganagic:1996pe,Cederwall:1996ri,Bergshoeff:1996tu,Aganagic:1996nn,Bergshoeff:1997kr,Skenderis:2002vf}. The matrix $\Gamma$ is determined by the kind and the configuration of the brane. For example, the kappa symmetry projection for D-branes in IIB theory can be expressed as
\begin{align}
 d^{p+1}\xi \Gamma=-e^{-\phi}(-\det(G+\Fcal))^{-1/2}e^{\Fcal}
 {\cal X}|_{(p+1)\text{-form}},\nn\\
 {\cal X}:=\sum_{n}\frac{1}{(2n)!}d\xi^{i_{2n}}\dots d\xi^{i_1}
 \Gamma_{\left\langle i_1\dots i_{2n}\right\rangle}\tau_3^{n}i\tau_2,\nn\\
 \Gamma_{\left\langle i_1\dots i_{s}\right\rangle}:=
\deldel{X^{m_1}}{\xi^{i_1}}\dots \deldel{X^{m_s}}{\xi^{i_s}}
 E^{a_1}_{m_1}\dots E^{a_s}_{m_s}\Gamma_{a_1\dots a_s},\label{kappa-D-brane}
\end{align}
where $\xi^{i}$ are the coordinates of the worldvolume, $X^{m}$ are the spacetime coordinates, and $E^a_m$ are the components of the vielbein defined by $E^{a}=E^{a}_{m}dX^{m}$. We also use in this expression the 2-form on the worldvolume $\Fcal=B+2\pi \alpha'F$, where $B$ is the pullback of the NSNS 2-form and $F$ is the field strength of the worldvolume gauge field.

Let us begin with $AdS_2 \times S^2$ D3-brane. This D3-brane extends to the $AdS_2\times S^2$ fiber direction. In order to preserve the SO(5) symmetry, this D-brane sits on a point in the base space where the $S^4$ shrinks, i.e. $\theta=0$ or $\theta=\pi$. We also have the electric field on the worldvolume which is proportional to the volume form of the $AdS_2$. The worldvolume coordinates are denoted by $\xi^0,\xi^1,\xi^4,\xi^5$ and the spacetime coordinate by $X^{0},X^1,X^4,X^5$. $X^0$ and $X^1$ parametrize the $AdS_2$ and $X^4,X^5$ parametrize $S^2$. We also assume the metric is diagonal for simplicity. The configuration of D3-brane we consider here is
\begin{align}
 X^0=\xi^0,\qquad X^1=\xi^1,\qquad X^4=\xi^4,\qquad X^5=\xi^5,\qquad 
u=\uk: \text{constant}.
\end{align}
The gauge field can be written as
\begin{align}
 \Fcal=\alpha E^{0}_{0}E^{1}_{1}d\xi^{0}d\xi^{1},
\end{align}
where $\alpha$ is a real constant. Here in this paper, we take the vielbein $E^{a}_{m}$ diagonal for simplicity\footnote{It is actually possible for the metric of eq.\eqref{metric1}.}. In this configuration of D3-brane, the kappa symmetry projection \eqref{kappa-D-brane}
\begin{align}
 \Gamma(D3)=\frac{1}{\sqrt{1-\alpha^2}}\left(
\Gamma_{0145}i\tau_2 + \alpha \Gamma_{45}\tau_1\right).
\end{align}
This operator acts on the ordinary IIB spinor. The corresponding operator acting on the $\epsilon$ spinor can be expressed as
\begin{align}
 \Gamma(D3)=\frac{1}{\sqrt{1-\alpha^2}}(-\mu_3\nu_3\tau_2+i\alpha\nu_3\tau_1).
\end{align}

For the D3-brane at $\theta=0$, the condition becomes $0=(1+\Gamma(D3))\epsilon$ with $\epsilon=\exp[(1/2)\mu_3\nu_3\tau_2 \uk]\zeta$ because of eq.\eqref{epsilon}. This condition leads to the following conditions for $\uk, \alpha$ and $\zeta$.
\begin{align}
 \sinh \uk=\frac{\sqrt{1-\alpha^2}}{|\alpha|},\qquad
\begin{cases}
  0=[1- \mu_3\tau_3]\zeta, & \alpha > 0\\
  0=[1+ \mu_3\tau_3]\zeta, & \alpha < 0  
\end{cases}.\label{susy3theta0}
\end{align}
Here $\alpha>0$ and $\alpha<0$ are two directions of electric field. They preserves opposite supersymmetry. Note that the projection conditions $0=[1\pm \mu_3\tau_3]\epsilon$ are compatible with the conditions \eqref{proj1},\eqref{weyl-proj}.

On the other hand, the D3-brane at $\theta=\pi$ preserves the following supersymmetry. The condition can be written as $0=(1+\Gamma(D3))\epsilon$ with $\epsilon=\exp[(1/2)\mu_3\nu_3\tau_2 \uk](-i\lambda_3\tau_2)\zeta$. This condition read the following conditions for $\uk$ and $\zeta$.
\begin{align}
 \sinh \uk=\frac{\sqrt{1-\alpha^2}}{|\alpha|},\qquad
\begin{cases}
  0=[1- \mu_3\tau_3]\zeta, & \alpha < 0\\
  0=[1+ \mu_3\tau_3]\zeta, & \alpha > 0  
\end{cases}.\label{susy3thetapi}
\end{align}
Note that the relation between the sign of $\alpha$ and the projection is opposite to the case of $\theta=0$.

\subsection{Supersymmetry of the $AdS_2\times S^4$ D5-branes}
Let us turn to the $AdS_2\times S^4$ D5-brane. In order to preserve the SO(3) symmetry, this D5-brane should sit on a point $u=0$ in the base space. We also have the gauge field excitation in the $AdS_2$ part. The worldvolume coordinates are denoted by $\xi^0,\xi^1,\xi^6,\dots,\xi^9$. The $S^4$ part of the spacetime coordinates are denoted by $X^6,\dots,X^9$. We consider the configuration of D5-brane expressed as\footnote{This configuration of D5-brane is a similar one as the ``baryon configuration'' in \cite{Imamura:1998gk}. Compared to \cite{Imamura:1998gk}, $\theta=$(constant) and the number of fundamental string charge is smaller than $N$ in the configuration of this paper.}
\begin{align}
 X^0=\xi^0,\qquad X^1=\xi^1,\qquad X^6=\xi^6,\dots , X^9=\xi^9,\nn\\
 u=0,\qquad \theta=\thet:\text{constant},\qquad
 \Fcal=\beta E^{0}_{0}E^{1}_{1}d\xi^{0}d\xi^{1}.
\end{align}
We obtain the kappa-symmetry projection for this D5-brane by inserting this configuration to \eqref{kappa-D-brane}. It can be written as
\begin{align}
 \Gamma(D5)=\frac{1}{\sqrt{1-\beta^2}}\left(
\Gamma_{016789}\tau_1-\beta\Gamma_{6789} i\tau_2
\right).
\end{align}
The operator that acts on $\epsilon$ spinor can be expressed as
\begin{align}
 \Gamma(D5)=\frac{1}{\sqrt{1-\beta^2}}(-\mu_3\lambda_3\tau_1-i\beta\lambda_3\tau_2).
\end{align}

The condition $0=(1+\Gamma(D5))\epsilon$ with $\epsilon=\exp\left(-\frac{i}{2}\lambda_3\tau_2\thet\right)\zeta$ reads the conditions on $\thet,\beta$ and $\zeta$. It can be written as
\begin{align}
 \cos\thet=\beta,\qquad
 0=[1-\mu_3\tau_3]\zeta.
\end{align}

As a result, the following three kinds of D-branes preserve the same half of the supersymmetry $0=[1-\mu_3\tau_3]\zeta$.
\begin{itemize}
 \item $AdS_2\times S^2$ D3-brane at $\theta=0$ with $\alpha>0$.
 \item $AdS_2\times S^2$ D3-brane at $\theta=\pi$ with $\alpha<0$.
 \item $AdS_2\times S^4$ D5-brane.
\end{itemize}

\subsection{String charges of AdS D-branes}

We expect that the $AdS_2\times S^4$ D5-brane corresponds to the anti-symmetric tensor representation. The rank of this anti-symmetric tensor is equal to the fundamental string charge of the D5-brane calculated in \cite{Pawelczyk:2000hy,Camino:2001at}. Let us review the result here. String charges can be expressed by the variation of the NSNS B-field.
\begin{align}
 k=2\pi \alpha' \frac{\delta S}{\delta B_{01}}=2\pi\alpha'\frac{\delta S}{\delta \Fcal_{01}}.
\end{align}

The bosonic part of the action of the D5-brane in this geometry can be written as
\begin{align}
 S_5=-T_5\int d^6\xi \sqrt{-\det(G+\Fcal)}+T_5\int \Fcal\wedge C_4,
\end{align}
where $T_5=1/((2\pi)^5\alpha'^3 g_s)$ is the tension of D5-brane. If we insert the configuration obtained in the previous subsection, the action becomes
\begin{align}
 S_5=&-T_5\int d^6 \xi E^0_0E^1_1e^6_6\dots e^9_9 L^4\sin^4\thet\sqrt{1-\beta^2}\nn\\
 &+T_5\int d^6 \xi \beta E^0_0E^1_1e^6_6\dots e^9_9 4L^4 \left(\frac38\thet
 -\frac14\sin2\thet+\frac{1}{32}\sin4\thet\right),
\end{align}
where we take the vielbein $E^{a}_{m}$ diagonal. Taking the definition of beta $\Fcal_{01}=\beta E^0_0 E^1_1$ into account, we can calculate the variation $\delta S_5/\delta \Fcal_{01}$. Then we insert the supersymmetry condition $\cos\thet=\beta$ into the result and we obtain the fundamental string charge of this D5-brane as \footnote{Here the volume of unit $S^4$ is $\int d\xi^6\dots d\xi^9\; e^{6}_{6}\dots e^{9}_{9}=\frac{8}{3}\pi^2$.}
\begin{align}
 k=\frac{2N}{\pi}\left(\frac12\thet-\frac14 \sin 2\thet\right).\label{D5k}
\end{align}
Note that this calculation of the string charge depends on the gauge of $C_4$. This ambiguity comes from the definition of the string charge in the presence of RR-flux. Here we employ the definition in which the single $AdS_2$ fundamental string at $u=\theta=0$ is $k=1$. In other words, we consider this fundamental string as the counterpart of the Wilson loop of the fundamental representation.
\section{$AdS_2\times S^4$ D5-brane solution and the circular Wilson loop}
\label{sec-circ}
In this section, we calculate the on-shell action of the $AdS_2\times S^4$ D5-brane. According to the AdS/CFT correspondence, this on-shell action is equal to the expectation value of the circular Wilson loop of anti-symmetric representation. We will show that it reproduces the same result as the matrix model calculation in a semi-classical approximation.

\subsection{Wick rotation of the solution and the on-shell action}
In order to consider the circular Wilson loops, we should perform Wick-rotation of the geometry. The metric of Euclidean version of $AdS_5$ can be written as
\begin{align}
 ds^2_{AdS_5}=\frac{L^2}{y^2}(dy^2+dr^2+r^2d\vp^2+dx_3^2+dx_4^2).
\end{align}
The $S^5$ part is the same as before. The RR 4-form is also Wick-rotated in the usual manner. The relevant part ($S^5$ part) of RR 4-form is the same as eq.\eqref{c4}.

We need to consider the action of the D5-brane. The bulk part of the Euclidean D5-brane action is written as
\begin{align}
 &S_{\bulk}=
T_{5}\int d^6\xi \sqrt{\det(G+\Fcal)}-iT_{5}\int \Fcal C_4.
\end{align}
We use the notation $S_{\bulk}$ because we will add certain boundary terms later. We can obtain the solution by Wick-rotation from the solution of the previous section. However since we need the conjugate momentums to calculate the boundary terms of the action, let us begin with the ansatz.

The ansatz used here is as follows. We identify one of spacetime coordinate $\vp$ as one of the worldvolume coordinate. The other worldvolume coordinate is denoted by $\rho$ and we assume $y$ and $r$ are functions of $\rho$. As in the previous section, we assume D5-brane is wrapped on a $S^4$ with $\theta=\thet$(constant). With this ansatz, the action can be written as
\begin{align}
 S_{\bulk}&=\int d\rho d\vp \Lcal_{\bulk},\\
   \Lcal_{bulk}&=T_5\frac{8}{3}\pi^2 L^4\left[ \sin^4\thet
\sqrt{\frac{L^4r^2}{y^4}(y'^2+r'^2)+\Fcal_{\rho\vp}^2}
-i\Fcal_{\rho\vp}\left(\frac32 \thet-\sin 2\thet+\frac18 \sin 4\thet\right)\right].
\end{align}
Here prime sign ``$'$'' denotes the $\rho$ derivative.

The conjugate momentum to $y$ becomes\footnote{Here ``conjugate momentum'' means the one when we consider $\rho$ as ``time.''}
\begin{align}
 p_y=\deldel{\Lcal_{bulk}}{y'}=T_5\frac 83 \pi^2 L^4 \sin^4 \thet
\frac{L^4r^2}{y^4}y'\left[\frac{L^4r^2}{y^4}(y'^2+r'^2)+\Fcal_{\rho\vp}^2\right]^{-1/2}.\label{py}
\end{align}
On the other hand, the conjugate momentum to $A_{\vp}$ can be written as
\begin{align}
 &p_{A}=(2\pi \alpha')\deldel{\Lcal_{\bulk}}{\Fcal_{\rho\vp}}\nn\\
&=(2\pi \alpha')T_5\frac 83 \pi^2 L^4\left[
\sin^4\thet
 \Fcal_{\rho\vp}\left[\frac{L^4r^2}{y^4}(y'^2+r'^2)+\Fcal_{\rho\vp}^2\right]^{-1
 /2}-i\left(\frac32 \thet -\sin 2\thet +\frac18 \sin 4\thet\right)\right].\label{pa}
\end{align}

Now let us turn to the solution. The solution can be written as
\begin{align}
 y=\rho,\qquad r=\sqrt{R^2-\rho^2},\qquad \Fcal_{\rho\vp}=-i\cos\thet \frac{L^2R}{\rho^2},
\end{align}
where $R$ is a real positive constant which correspond to the radius of the circular Wilson loop. What we want to do is to insert this solution to the action and obtain on-shell action. When we insert this solution in to $S_{\bulk}$, the integral is divergent near $\rho=0$. In order to treat this quantity, we introduce a cut off $\rho_0$. Then the on-shell value of $S_{\bulk}$ inserted this solution becomes
\begin{align}
 S_{\bulk}&=\int_{\rho_0}^{R}d\rho \int_{0}^{2\pi}d\vp \Lcal_{\bulk}\nn\\
&=\left(-1+\frac{R}{\rho_0}\right)\frac{2N}{3\pi}\sqrt{\lambda}
\left[\sin^5\thet -\cos\thet\left(\frac32\thet-\sin 2\thet+\frac18\sin 4\thet\right)\right].
\end{align}
In order to treat the effect of the boundary correctly, we should introduce the boundary terms. Here we follow the procedure of \cite{Drukker:2005kx,Drukker:1999zq}. We should solve the equation of motion with the boundary condition $\delta p_{y}=0,\ \delta p_{A}=0$ instead of $\delta y=0,\ \delta A$=0. To realize this boundary condition from the variation of the action, we should add the following boundary terms to the action.
\begin{align}
 S_{\bdy,y}&=-\left. \int_{0}^{2\pi}d\vp p_y  y \right|_{\rho=\rho_0} =\left(-\frac{R}{\rho_0}+\frac{\rho_0}{R}\right) \frac{2N}{3\pi}\sqrt{\lambda} \sin^3\thet,\\
 S_{\bdy,A}&=-\int_{\rho_0}^{R}d\rho \int_{0}^{2\pi}d\vp\; p_{A}
 \frac{1}{2\pi \alpha'} \Fcal_{\rho\vp}\nn\\
 &=\left(1-\frac{R}{\rho_0}\right)\frac{2N}{3\pi}\sqrt{\lambda}
\left[-\sin^3\thet+\sin^5\thet-\cos\thet\left(\frac 32\thet -\sin 2\thet +\frac 18 \sin4\thet\right)\right].
\end{align}

If one sums up these three contributions, the divergence cancels and in the limit $\rho_0\to 0$ the total action becomes
\begin{align}
S_{\rm tot}= S_{\bulk}+S_{\bdy,y}+S_{\bdy,A}=-\sqrt{\lambda}\frac{2N}{3\pi}\sin^3\thet.\label{stot}
\end{align}
This is the final result of the calculation of AdS side. The expectation value of the Wilson loop is $\exp(-S_{\rm tot})$ in the calculation in AdS side. Note that this result is equal to the naive calculation using a string in $AdS_5$ with effective tension $1/(2\pi \alpha') 2N/(3 \pi) \sin^3\thet$ of \cite{Pawelczyk:2000hy,Camino:2001at}.

Let us here do some simple check. The string charge $k$ is related to $\thet$ as eq. \eqref{D5k}. When $k$ is much smaller than $N$, the angle $\thet$ becomes small. In this case, k is related to $\thet$ by
\begin{align}
 k\cong \frac{2N}{3\pi}\thet^3.
\end{align}
If we insert this expression to eq.\eqref{stot} and take $\thet \ll 1$ into account, we obtain
\begin{align}
 S_{\rm tot}\cong -k\sqrt{\lambda}.
\end{align}
This is the expected one from $k$ overwrapping fundamental strings. Actually our result includes the corrections of $k/N$. The first few terms look like
\begin{align}
 S_{\rm tot}=-k\sqrt{\lambda}\left[
1-\frac{3}{10}\left(\frac{3\pi k}{2N}\right)^{2/3}
-\frac{3}{280}\left(\frac{3\pi k}{2N}\right)^{4/3}
-\frac{11}{8400}\left(\frac{3\pi k}{2N}\right)^{2}
+O\left(\frac{k}{N}\right)^{8/3}
\right].
\end{align}
Interestingly, these corrections includes the fractional power of $k/N$.

Another check is related to the complex conjugation. The rank $k$ anti-symmetric representation of SU($N$) and the rank $(N-k)$ anti-symmetric representation are related by complex conjugation. We expect that the VEV of these Wilson loops are the same. Actually the parameter $\thet$ and $\theta_{(N-k)}$ are related as $\theta_{(N-k)}=\pi-\theta_k$ because of eq.\eqref{D5k}. Therefore $S_{\rm tot}$ in \eqref{stot} for $k$ and $(N-k)$ are the same.
\subsection{Gaussian matrix model calculation}

It is conjectured that the vacuum expectation value of the circular Wilson loop in $\Ncal=4$ U($N$) Yang-Mills theory can be calculated by the Gaussian matrix model \cite{Erickson:2000af,Drukker:2000rr}. In this subsection, we calculate the expectation value of the Wilson loop of anti-symmetric representation by the Gaussian matrix model in semi-classical approximation.

Let us consider the one matrix model with Gaussian potential. The expectation value of a gauge invariant function $f(M)$ of the $N\times N$ hermitian matrix $M$ is defined as
\begin{align}
 &\mmev{f(M)}:=\frac{1}{Z}\int dM f(M)\exp\left(-\frac{2N}{\lambda} \tr [M^2]\right),\\
 &Z:=\int dM \exp \left(-\frac{2N}{\lambda} \tr [M^2]\right).
\end{align}
Here ``gauge invariant'' means $f$ satisfies $f(VMV^{-1})=f(M)$ for any unitary matrix $V$. In the above definition, the measure $dM$ is $dM=\prod dM_{ij}$ with the constraint $M^{\dag}=M$.

The expectation value of the circular Wilson loop $\Tr_{R}[U]$, $U:=P \exp \oint ds[i A_{\mu}\dot{x}^{\mu}+ \vp_4 |\dot{x}|]$ in $\Ncal=4$ Yang-Mills theory can be calculated by this matrix model as
\begin{align}
 \left\langle \Tr_{R}[U] \right\rangle_{\rm YM}=\mmev{\Tr_{R}[e^{M}]}.
\end{align}

One of the standard methods to evaluate the matrix integral is diagonalizing the matrix.  For example, the partition function $Z$ can be rewritten as the integral of the eigenvalues $m_1, \dots, m_N$ as
\begin{align}
 Z&=C_N\int \prod_{j=1}^{N} dm_j \prod_{1\le i <j\le N}(m_i-m_j)^2
 \exp\left(-\frac{2N}{\lambda} \sum_{j=1}^{N}m_j^2\right)\nn\\
 &=C_N\int \prod_{j=1}^{N} dm_j \exp(-S_{e}(m)),\\
 S_{e}(m)&:=\frac{2N}{\lambda}\sum_{j=1}^{N}m_j^2
 -2\sum_{i<j}\log|m_i-m_j|,
\end{align}
where $C_N$ is a constant which comes from the Jacobian. Let us consider the saddle point of this theory. The ``equation of motion'' for $m_i$ becomes 
\begin{align}
0= \frac{4N}{\lambda}m_i-2\sum_{j\ne i}\frac{1}{m_i-m_j}.
\end{align}
It is convenient to introduce the resolvent $\omega(z):=\tr \frac{1}{M-z}=\sum_{j=1}^{N}\frac{1}{m_j-z}$. At the saddle point, we can derive the following differential equation for this resolvent from the equations of motion.
\begin{align}
 0=\frac{4N^2}{\lambda}+\frac{4N}{\lambda}z\omega(z)+\omega(z)^2
 -\omega'(z).
\end{align}
In the large $N$ limit, the last term $\omega'(z)$ can be negligible. In this case, the above differential equation becomes a algebraic equation and is easily solved as
\begin{align}
 \omega(z)=2N\left(-\frac{z}{\lambda}\pm\sqrt{\left(\frac{z}{\lambda}\right)^2
-\frac{1}{\lambda}}\right).
\end{align}
The density of eigenvalues is expressed as
\begin{align}
 \rho(x)=\frac{1}{2\pi i} (\omega(x-i\epsilon)-\omega(x+i\epsilon))
  =\frac{2N}{\pi\lambda}\sqrt{\lambda-x^2}.
\end{align}
Let $\mcl_j(0),\ j=1,\dots,N$ be the eigenvalues of the solution of the classical equation of motion with the order $\mcl_1(0)>\mcl_2(0)>\mcl_3(0)>\dots >\mcl_N(0)$. Also let us define $\thet$ by the classical value $\mcl_{k}(0)=:\sqrt{\lambda}\cos \thet$. Since $k$ is the number of eigenvalues between $\mcl_1(0)=\sqrt{\lambda}$ and $\mcl_k(0)=\sqrt{\lambda}\cos \thet$, we can write down the relation between $k$ and $\thet$ using the above eigenvalue density $\rho(x)$ as
\begin{align}
 k=\int_{\sqrt{\lambda}\cos \thet}^{\sqrt{\lambda}}dx \; \rho(x)
=\frac{2N}{\pi}\left(\frac12\thet-\frac14\sin 2\thet\right). \label{changetotheta}
\end{align}
This is consistent with the equation \eqref{D5k}.

Now let us consider the rank $k$ anti-symmetric representation $A_{k}$. The VEV of the Wilson loop $\Tr_{A_k}U$ can be calculated by the matrix integral
\begin{align}
 \mmev{\Tr_{A_k}[e^M]}=\frac{1}{Z}\int dM \Tr_{A_k}[e^M] \exp\left(-\frac{2N}{\lambda} \tr [M^2]\right).
\end{align}
When we diagonalize the matrix $M$, the trace of anti-symmetric representation becomes
\begin{align}
 \Tr_{A_k}[e^M]=\sum_{1\le j_1 < j_2<\dots <j_k\le N}\exp\left(
m_{j_1}+\dots + m_{j_k}\right).
\end{align}
In the integral of the eigenvalues, all the eigenvalues are equivalent. So the expectation values of all terms in the above sum become the same. Therefore we can write the expectation value of the $\Tr_{A_k}[e^M]$ as\footnote{Note that in \cite{Drukker:2005kx} the Wilson loop is multiply winded fundamental one and the operator evaluated in matrix model is the operator $\tr[e^{kM}]=\sum_{j=1}^{N}e^{km_j}$. So the operator in the integrand of the eigenvalue integral is $e^{km_1}$. This is the crucial difference between the matrix model calculation of \cite{Drukker:2005kx} and ours in this paper.}
\begin{align}
 \mmev{\Tr_{A_k}[e^M]}
 =\frac{C_{N}}{Z}\binom{N}{k}\int \prod_{j=1}^{N} dm_j \; \exp(m_1+\dots +m_k) \exp(-S_{e}(m)).
\label{aint}
\end{align}
The insertion of the operator looks like a source term for the theory. In order to evaluate this integral, let us consider the saddle point of the following action including the source.
\begin{align}
 S(k,m)&:=S_{e}(m)-\sum_{i=1}^{k}m_i\nn\\
&=\frac{2N}{\lambda}\sum_{j=1}^{N}m_j^2
 -2\sum_{i<j}\log|m_i-m_j|-\sum_{i=1}^{k}m_i.
\end{align}
We would like to evaluate the integral by saddle point. The classical equation motion can be written as
\begin{align}
 \deldel{S}{m_j}=0.
\end{align}
Let us write the solution of this equation as $\mcl_j(k)$. We put the argument $k$ in order to remember that it depends on $k$. The on-shell action can be written as
\begin{align}
 \Scal(k):=S(k,\mcl(k))-S(0,\mcl(0)).
\end{align}
We can approximate the integral \eqref{aint} by
\begin{align}
 \mmev{\Tr_{A_k}[e^M]}
 \cong \exp(-\Scal(k)).
\end{align}
Let us regard $k$ as a continuous variable and differentiate $\Scal$ by $k$.
\begin{align}
 \deldel{\Scal}{k}=\sum_{j=1}^N \left(\deldel{\mcl_j(k)}{k}\right)
\deldel{S}{m_j}(k,\mcl(k))+\deldel{S}{k}(k,\mcl(k)).
\end{align}
The first sum vanishes because of the equation of motion. As for the second term, the differential can be written as $S(k+1,m)-S(k,m)=-m_k$. We can also replace $\mcl_k(k)$ by $\mcl_k(0)$ because the difference is very small, namely the relative error $(\mcl_k(k)-\mcl_k(0))/\mcl_k(0)\sim \sqrt{\lambda}/N \ll 1$. As a result, we can write the derivative of $\Scal$ as
\begin{align}
 \deldel{\Scal}{k}=-\mcl_{k}(0).
\end{align}
The derivative $\Scal$ by the parameter $\thet$ of eq. \eqref{changetotheta} can be expressed as
\begin{align}
 \deldel{\Scal}{\thet}=\deldel{k}{\thet}\deldel{\Scal}{k}=-\frac{2N\sqrt{\lambda}}{\pi}\sin^2\thet\cos\thet.
\end{align}
This equation can be easily integrated with the boundary condition $\Scal(k=0)=0$ as
\begin{align}
 \Scal=-\frac{2N\sqrt{\lambda}}{3\pi}\sin^3\thet.
\end{align}
This is completely the same as \eqref{stot}.

\subsection*{Acknowledgment}
I would like to thank Mohab Abou-Zeid, Mitsuhiro Arikawa, Tohru Eguchi, Nadav Drukker, Bartomeu Fiol, Kazuo Hosomichi, Yosuke Imamura, So Matsuura, Sanefumi Moriyama, Nikita Nekrasov, Kazutoshi Ohta, Jeong-Hyuck Park, Gordon W. Semenoff, Tadashi Takayanagi, Jan Troost, Tatsuya Tokunaga, Marcel Vonk, and Shing-Tung Yau for useful discussions and comments. I am also grateful for the hospitality of Theoretical Physics Group of RIKEN, Yukawa Institute of Theoretical Physics, Theoretical Group of KEK, and Niels Bohr Institute. This work was supported in part by the European Research Training Network contract 005104 ``ForcesUniverse.''

\appendix
\section{Spinors in $AdS_2\times S^2 \times S^4$ fibration}
\label{app1}
Since in $AdS_5\times S^5$ background only the metric and the RR 5-form $G_5$ are excited, the invariance of the gravitino under supersymmetry transformation can be written as
\begin{align}
 &\delta \psi_{M}=\nabla_{M}\eta+\frac{i}{2}\sla{G}_5\tau_2 \Gamma_{M}\eta.
\end{align}
The dilatino condition becomes trivial. The parameter of supersymmetry is a doublet of Majorana-Weyl spinors. We use the Pauli matrices $\tau_j$ to rotate this doublet.

We use the 10-dimensional gamma matrices as follows.
\begin{align}
 &\Gamma^0=\sigmac^{0}\otimes \sigc \otimes 1 \otimes 1,
 &&\Gamma^1=\sigmac^{1}\otimes \sigc \otimes 1 \otimes 1,\nn\\
 &\Gamma^2=1\otimes \sigma_1\otimes 1 \otimes 1,
 &&\Gamma^3=1\otimes \sigma_2\otimes 1 \otimes 1,\nn\\
 &\Gamma^4=\sigmac^{3}\otimes \sigc \otimes \sigmah_4 \otimes 1,
 &&\Gamma^5=\sigmac^{3}\otimes \sigc \otimes \sigmah_5 \otimes 1,\nn\\
 &\Gamma^a=\sigmac^{3}\otimes \sigc \otimes \sigmah_6 \otimes \gamma^{a},
&&\qquad (a=6,7,8,9),
\end{align}
where $(\sigma_1,\sigma_2,\sigc)$ and $(\sigmah_4,\sigmah_5,\sigmah_6)$ are sets of the Pauli matrices. $(\sigmac_1,\sigmac_2,\sigmac_3)$ is another set of the Pauli matrices, and we defined $\sigmac^{0}$ as $\sigmac^{0}:=i\sigmac_2$. $\gamma^{a}, \ (a=6,7,8,9)$ are gamma matrices of Euclidean 4 dimensions.

\begin{align}
 &\nablac_p \chic^{I}_{a}=\frac i2 a \sigmac_{p}\chic^{I}_{-a},\qquad 
 \sigmac_{3}\chic^{I}_{a}=a\chic^{I}_{a},\qquad
  (p=0,1,\qquad a=\pm 1,\qquad I=1,2),\\
  &\nablac_p \chih^{J}_{b}=\frac 12 b \sigmah_{p}\chih^{J}_{-b},\qquad 
 \sigmah_{6}\chih^{J}_{b}=b\chih^{J}_{b},\qquad
 (p=4,5,\qquad b=\pm 1,\qquad J=1,2),\\
  &\nablac_p \chi^{K}_{c}=\frac 12 c \gamma_{p}\chi^{K}_{-c},\qquad 
 \gamma_{6789}\chi^{K}_{c}=c\chi^{K}_{c},\qquad
 (p=6,\dots,9,\qquad b=\pm 1,\qquad K=1,2,3,4).
\end{align}
where $\nablac$ is the covariant derivative of the Levi-Civit\'a connection of the unit $AdS_2$, $S^2$ or $S^4$.

One can reduce the problem to 2-dimensions by expanding the 10-dimensional spinor pair $\eta$ by the above Killing spinors.
\begin{align}
 \eta=\sum_{a,b,c}\chic^{I}_{a}\otimes\epsilon_{abcIJK}\otimes\chih^{J}_{b}
\otimes \chi^{K}_{c}.
\end{align}
Each $\epsilon_{abcIJK}$ is a pair of 2-dimensional spinors. $\Gamma_2,\Gamma_3,\sigc,\tau_1,\tau_2,\tau_3$ act on $\epsilon_{abcIJK}$.

Let $\mu_j,\nu_j,\lambda_j,\ (j=1,2,3)$ be sets of the Pauli matrices which act on the indices $a,b,c$ respectively. This means, for example,
\begin{align}
 (\mu_j \epsilon)_{abc}=(\mu_j)_{aa'}\epsilon_{a'bc}.
\end{align}
We also define a set of matrices $\rho_j,\ (j=1,2,3)$ as
\begin{align}
 \rho_1=\mu_3\tau_3,\qquad \rho_2=\nu_3\tau_1,\qquad \rho_3=\mu_3\nu_3\tau_2.
\end{align}
These three matrices $\rho_j$ satisfy the algebra of the Pauli matrices.

\providecommand{\href}[2]{#2}\begingroup\raggedright\endgroup

\end{document}